\documentclass[amssymb,prl,twocolumn,notitlepage]{revtex4-1}

\renewcommand{\vec}[1]{\mathbf{#1}}
\newcommand{\uvec}[1]{\vec{\widehat{#1}}}

\newcommand{\braket}[2]{\langle #1 | #2 \rangle}
\newcommand{\kets}[1]{| #1 \rangle}

\newcommand{\ket}[1]{| #1 \rangle}
\newcommand{\bra}[1]{\langle #1 |}

\makeatletter
\def\Ddots{\mathinner{\mkern1mu\raise\p@
\vbox{\kern7\p@\hbox{.}}\mkern1mu
\raise4.5\p@\hbox{.}\mkern1mu\raise7.5\p@\hbox{.}\mkern1mu}}
\makeatother

\makeatletter
\def\DDdots{\mathinner{\mkern1mu\raise7.5\p@
\vbox{\kern7\p@\hbox{.}}\mkern1mu
\raise4.5\p@\hbox{.}\mkern1mu\raise\p@\hbox{.}\mkern1mu}}
\makeatother

\usepackage{color}

\usepackage{amsmath}                          
\usepackage{amsfonts}                         
\usepackage{amssymb}                          
\usepackage{amsthm}                           
\usepackage{mathdots}                         

\usepackage{graphicx}                         
\usepackage{subfigure}                        
\usepackage{overpic}

\usepackage{datetime}                         
\usepackage{comment}                          
\usepackage{acronym}			
\usepackage{array}
\usepackage{fancyvrb}

\theoremstyle{plain}

\newacro{QLSA}[QLSA]{Quantum Linear Systems Algorithm}
\newacro{AE}[AE]{Amplitude Estimation}
\newacro{FEM}[FEM]{Finite Element Method}
\newacro{UTD}[UTD]{Uniform Theory of Diffraction}
\newacro{SPAI}[SPAI]{sparse approximate inverse}

\usepackage[showonlyrefs]{mathtools}

\begin{document}


\title{Preconditioned quantum linear system algorithm}
\author{B. D. Clader}\email{dave.clader@jhuapl.edu}
\author{B. C. Jacobs}
\author{C. R. Sprouse}
\affiliation{The Johns Hopkins University Applied Physics Laboratory, Laurel, MD 20723, USA}

\begin{abstract}
We describe a quantum algorithm that generalizes the quantum linear system algorithm [Harrow {\it et al}., Phys. Rev. Lett. {\bf 103}, 150502 (2009)] to arbitrary problem specifications. We develop a state preparation routine that can initialize generic states, show how simple ancilla measurements can be used to calculate many quantities of interest, and integrate a quantum-compatible preconditioner that greatly expands the number of problems that can achieve exponential speedup over classical linear systems solvers.  To demonstrate the algorithm's applicability, we show how it can be used to compute the electromagnetic scattering cross section of an arbitrary target exponentially faster than the best classical algorithm.
\end{abstract}

\maketitle

The potential power of quantum computing was first described by Feynman, who showed that the exponential growth of the Hilbert space of a quantum computer allows efficient simulations of quantum systems, whereas a classical computer would be overwhelmed \cite{springerlink:10.1007/BF02650179}.  Shor extended the applicability of quantum computing when he developed a quantum factorization algorithm that also provides exponential speedup over the best classical algorithm \cite{Shor1994}. 

More recently, Harrow \textit{et al.} \cite{PhysRevLett.103.150502} demonstrated a quantum algorithm for solving a linear system of equations that, for well-conditioned matrices, gives exponential speedup over the best classical method.  In that paper, the authors demonstrated how to invert a sparse matrix to solve the quantum linear system $A\ket{x}=\ket{b}$.  The requirements for achieving exponential speedup were: 1) the elements of $A$ be efficiently computable via a black-box oracle; 2) the matrix $A$ must to be sparse, or efficiently decomposable into sparse form; 3) the condition number of $A$ must scale as polylog $N$ where $N$ is the size of the linear system.

As presented, the algorithm had three features that made it difficult to apply to generic problem specifications and achieve the promised exponential speedup.  These included: {\em State preparation}  - preparing the generic state $\ket{b}$ is an unsolved problem \cite{Grover02, Aharonov:2003:AQS:780542.780546, ward:194105, PhysRevA.73.012307, Kaye04}, and no mention on how one might do this was provided.   {\em Solution readout} - since the solution is stored in a quantum state $\ket{x}$, measurement of it is impractical. The authors suggested that it could be used to calculate some expectation values of an arbitrary operator $\bra{x} \hat{R} \ket{x}$.  However, no measurement procedure was specified, and estimating $\bra{x} \hat{R} \ket{x}$ is not trivial in general.  {\em Condition number} - in order for the quantum algorithm to achieve exponential speedup, the condition number can scale at most poly logarithmically with the size of the matrix $A$.  This is a very strict condition that greatly limits the class of problems that can achieve exponential speedup.  

In this letter, we provide solutions to these three problems, greatly expanding the applicability of the \ac{QLSA}.  In addition, we show how our new techniques enable the first start-to-finish application of the \ac{QLSA} to a problem of broad interest and importance.  Namely, we show how to solve for the scattering cross section of an arbitrary target exponentially faster than the best classical algorithm.

Before we begin, we first review the original scheme of Harrow \textit{et al.} \cite{PhysRevLett.103.150502}.  One begins by preparing a quantum state $\ket{\Psi} = \sum_{\tau=0}^{N-1}\ket{\tau}\ket{b}$.   Next, perform a phase-estimation routine by simulating the matrix $A$ as a Hamiltonian for time $\ket{\tau}$ giving
\begin{equation}
\label{eq:simpleHamSim}
\ket{\Psi} \to  \sum_{j=0}^{N-1}\sum_{\tau=0}^{T-1}\ket{\tau}e^{i\lambda_j\tau t_0/T}\beta_j\ket{u_j},
\end{equation}
with $t_0=O(\kappa/\epsilon)$, where $\kappa$ is the condition number of the matrix $A$, and $\epsilon$ is the desired numerical accuracy of the algorithm.  This scaling is determined by error requirements, and implies that the total quantum algorithm complexity scales linearly with $\kappa$.  To obtain Eq. \eqref{eq:simpleHamSim} we have expanded the state-vector $\ket{b}$ in the eigen-basis of $A$ with eigenvalues $\lambda_j$ and eigenvalues $\ket{u_j}$.  Apply a quantum Fourier transform to the first register yielding
\begin{equation}
\label{eq:simpleQFT}
\ket{\Psi} \to \sum_{j=0}^{N-1}\ket{\tilde{\lambda}_j}\beta_j\ket{j},
\end{equation}
where $\tilde\lambda_j$ is related to the eigenvalues of $A$ through a constant scaling.  Apply a rotation to an adjoined ancilla qubit, controlled off the value of the first register yielding
\begin{equation}
\label{eq:simpleSolution}
\ket{\Psi} \to \sum_{j=0}^{N-1}\ket{\tilde{\lambda}_j}\beta_j\ket{j}\left(\sqrt{1-\frac{C}{\lambda_j}}\ket{0}_a + \frac{C}{\lambda_j}\ket{1}_a\right),
\end{equation}
where $C$ is a normalization constant chosen to ensure rotations are less than $2\pi$, and the ancilla qubit is denoted by the subscript $a$.  One then uncomputes the first register by reversing the previous steps and measures the ancilla qubit.  If the measurement result is $\ket{1}$ we obtain
\begin{equation}
\label{eq:finalSolution}
\ket{\Psi} \to C^\prime\sum_{j=0}^{N-1}\frac{\beta_j}{\lambda_j}\ket{j} \equiv \ket{x},
\end{equation}
the solution to $A\ket{x} = \ket{b}$, with normalization factor $C^\prime$.  

With this starting point, we present robust approaches to issues highlighted regarding state-preparation, solution read-out, and condition number that are not addressed in the scheme outlined in Eqs. \eqref{eq:simpleHamSim} - \eqref{eq:finalSolution}.  Direct preparation of the state $\ket{b}$, required by Eq. \eqref{eq:simpleHamSim}, is not possible in general.  Consider instead the state
\begin{equation}
\label{initial_state}
\ket{b_T} = \cos{\phi_b}\kets{\tilde{b}}\ket{0}_a +  \sin{\phi_b}\ket{b}\ket{1}_a
\end{equation}
that contains our desired arbitrary state, $\ket{b}$, entangled with an ancilla qubit in state $\ket{1}_a$.  This can be prepared efficiently in the following manner:  initialize three quantum registers and an ancilla qubit as
\begin{equation}
\label{initial_state_stateprep}
\ket{\Psi} = \frac{1}{\sqrt{N}}\sum_{j=0}^{N-1}\ket{j} \ket{0} \ket{0} \ket{0}_a.
\end{equation}
Query a black-box oracle function that calculates the amplitude and phase components, denoted as $b_j$ and $\phi_j$ respectively, of the vector $\ket{b}=\sum_{j=0}^{N-1}b_je^{i\phi_j}\ket{j}$, controlled off the value in the first register.  Apply a controlled phase gate to the ancilla qubit, controlled by the calculated value of the phase, and finally rotate the fourth ancilla qubit controlled by the calculated value of the amplitude.  Uncompute registers 2 and 3 by calling the $b_j$ and $\phi_j$ oracle function again leaving
\begin{equation}
\label{state_4_stateprep}
\ket{\Psi} \to \frac{1}{\sqrt{N}}\sum_{j=0}^{N-1}e^{i\phi_j} \ket{j}\left(\sqrt{1-C_b^2 b_j^2}\ket{0}_a + C_b b_j\ket{1}_a\right),
\end{equation}
where $C_b \le 1/{\rm max }(b_j)$ to ensure that all rotations are less than $2\pi$.  State \eqref{state_4_stateprep} is exactly the state \eqref{initial_state} with $\sin^2 \phi_b = \frac{C_b^2}{N}\sum_{j=0}^{N-1}b_j^2$, $\cos^2 \phi_b = \frac{1}{N}\sum_{j=0}^{N-1}(1-C_b^2b_j^2)$, $\ket{\tilde b} = \frac{1}{\sqrt{N}\cos\phi_b}\sum_{j}\sqrt{1-C_b^2b_j^2}e^{i\phi_j}\ket{j}$, and $\ket{b} = \frac{C_b}{\sqrt{N}\sin\phi_b}\sum_{j}b_je^{i\phi_j}\ket{j}$.  This state can be prepared efficiently so long as the oracle used to compute $b_j$ and $\phi_j$ is efficient.

Next, we apply the Eqs. \eqref{eq:simpleHamSim} to \eqref{eq:simpleSolution} of the original \ac{QLSA} to the state $\ket{b_T}$. We modify the original algorithm by removing the last post-selection step in Eq. \eqref{eq:finalSolution}, such that the our implementation is unitary.  This yields
\begin{equation}
\label{eq:solution}
\ket{\Psi} = (1-\sin^2{\phi_b}\sin^2{\phi_x})^{1/2}\ket{\Phi_0} + \sin{\phi_b}\sin{\phi_x}\ket{x} \ket{1}_a\ket{1}_a,
\end{equation}
where $\sin\phi_x$ is a normalization term resulting from the \ac{QLSA} and related to $C$ in Eq. \eqref{eq:simpleSolution}, $\ket{\Phi_0}$ is a garbage state in an expanded Hilbert space spanned by the solution vector and two ancilla qubits which are not in the state $\ket{1}$ simultaneously, and $\ket{x}$ is the normalized solution to the linear systems problem entangled with two ancilla qubits in the state $\ket{1}_a$.  As shown, Eq. \eqref{eq:solution} contains the solution $\ket{x}=A^{-1}\ket{b}$ for an arbitrary input state $\ket{b}$.

We now provide  a resolution to the read-out problem, and show how to unentangle the solution $\ket{x}$ from the rest of state \eqref{eq:solution}.  While access to the entire solution is impossible since it lies in an exponentially large space, we provide three examples of calculable quantities from Eq. \eqref{eq:solution}.  These are: the overlap of the solution with an arbitrary vector $\ket{R}$, moments of the solution $\bra{x}x^n\ket{x}$, as well as individual values of the solution vector denoted $x_j = \braket{j}{x}$.  

To estimate the overlap, we prepare the state $\ket{R_T} = \cos{\phi_r} \kets{\tilde{R}}\ket{0} + \sin{\phi_r}\kets{R}\ket{1}$ using the same method we used to prepare the state $\kets{b_T}$.   Adjoin this state to Eq. \eqref{eq:solution} along with a fourth ancilla qubit initialized to state $\ket{0}_a$.  Apply a Hadamard gate to the fourth ancilla qubit, and use it to perform a controlled swap operation between the registers containing the solution vector $\ket{x}$ and the vector $\ket{R}$, followed by a second Hadamard operation on the ancilla.  In doing so, we compute the dot product between $\ket{x}$ and $\ket{R}$ as
\begin{equation}
\label{overlap}
|\braket{R}{x}|^2 =  \frac{P_{1110} - P_{1111}}{\sin^2{\phi_b}\sin^2{\phi_x}\sin^2{\phi_r}},
\end{equation}
where $P_{1110}$ and $P_{1111}$ refer to the probability of measuring a 1 in the first three ancilla qubits, and a 0 or 1 in the last adjoined ancilla respectively. 

One can use a slightly modified version of Eq. \eqref{overlap} to calculate moments of the solution $\bra{x}x^n\ket{x}$.  One does this by applying a rotation to an ancilla, controlled by the state $\ket{x}$ with the operator $H_{rw} = x^n\ket{x}\bra{x}$.  Taking $\ket{R}=\ket{x}$ allows us to compute any moment of the solution using Eq. \eqref{overlap} with one additional ancilla measurement.  To calculate a particular solution value, one uses \ac{AE} \cite{Brassard00} targeted at the desired $j$ value.  In this manner, Eq. \eqref{eq:solution} together with \ac{AE} can estimate any $x_j = \braket{j}{x}$ efficiently.  All of these results are obtainable as they only require ancilla measurements.  Because the techniques used for state preparation and linear system solving are unitary, the various amplitudes can be estimated using \ac{AE}.  We anticipate this generalized procedure being useful for other algorithms that use the \ac{QLSA} such as the quantum data-fitting algorithm \cite{PhysRevLett.109.050505}.

The last, and most critical issue, relates to the spectral condition number $\kappa$.  The Hamiltonian simulation step in Eq. \eqref{eq:simpleHamSim}, causes the quantum algorithm query complexity to scale linearly with $\kappa$.  Thus, in order for the quantum algorithm to scale as $O(\log N)$ and achieve exponential speedup, $\kappa$ must scale in the worst case poly logarithmically with the size of the $N\times N$ matrix $A$.  However for most matrices one typically has linear or even exponential scaling with $N$ \cite{Brenner2008,bank1989}, greatly limiting the class of problems that can achieve exponential speedup.

We provide a solution to the condition number scaling problem through a technique known as preconditioning \cite{benzi2002preconditioning}.  When preconditioning, rather than solving the system $Ax=b$, one instead solves the modified linear system $MAx=Mb$.  Convergence is improved if one can find a matrix $M$ such that the condition number of $MA$ is much lower than the original matrix $A$.  The best preconditioner is obviously $M=A^{-1}$.  However, finding $A^{-1}$ is equivalent to solving the linear system, so using this as  a preconditioner provides no speedup.  One solution is to find an efficiently computable approximate inverse $M\approx A^{-1}$.  Unfortunately, two constraints imposed by the quantum algorithm make many classical preconditioners unusable.  These two constraints are 1) only local knowledge of $A$ can be obtained in order for the algorithm to be efficient, and 2) the preconditioned matrix $MA$ must itself be sparse for Hamiltonian simulation.  

 A class of preconditioners that satisfy both these constraints are \ac{SPAI} preconditioners \cite{doi:10.1137/S1064827594276552, doi:10.1137/S106482759833913X}.  We integrate this method with the quantum algorithm as follows.  One attempts to find the matrix $M$ by minimizing
 \begin{equation}
\label{eq:spaiminfunc}
||MA-I||_F^2 = \sum_{k=0}^{N-1}||(MA-I)e_k||_2^2,
\end{equation}
where the subscript $F$ refers to the Frobenius norm, and $e_k = (0,\cdots,0,1,0,\cdots,0)^T$.  Eq. \eqref{eq:spaiminfunc} separates into $N$ independent least squares problems
\begin{equation}
\label{eq:indepspai}
\min_{\hat{m}_k} ||\hat{A} \hat{m}_k - \hat{e}_k ||_2
\end{equation}
for $k=0,\cdots,N-1$, where the circumflex denotes the subspace of only non-zero matrix vector products.  One imposes sparsity constraints on matrix $M$.  Therefore the least squares problem in Eq. \eqref{eq:indepspai} is very small, of order $n \times d$ where $n$ is the number of non-zero rows in column $k$ and $d$ is the number of nonzero elements per row.  Thus, we now have $N$ independent $n \times d$ sized least squares problems to compute the \ac{SPAI} preconditioner.  We have a black-box oracle function that can compute the elements and locations of non-zero terms in the matrix $A$ for a given row.  Therefore calls this oracle to setup Eq. \eqref{eq:indepspai} controlled by a supplied row index.  Since the matrix $A$ is highly sparse, both $n$ and $d$ are small.

Within the quantum algorithm, to simulate the matrix $A$, one requires a unitary $U^{(c)}$ that calculates the element of $MA$ and its column index $y_k$ for a specific graph edge color $c$ (see supplementary material and Refs. \cite{aharonov2003adiabatic, Childs:2003:EAS:780542.780552, 1751-8121-44-44-445308} for more information), conditioned on a row index $k$.  This operates as $U^{(c)}\ket{k,0} = \ket{k,a_k,y_k}$.  The matrix preconditioner step can fit neatly within this unitary operator.  The techniques used to calculate the \ac{SPAI} require only local accesses of $A$, which we have access to via the oracle for $A$, and the matrix $M$ can be calculated for each row independently.  The sparsity structure of $M$ is either calculated efficiently or set a priori, and thus we can calculate $y_k$ efficiently.  Therefore, the oracle for the matrix $MA$ can be created by combining Eq. \eqref{eq:indepspai} together with the original oracle for $A$.  

To create the state $M\ket{b}$ is similar.  In Eqs. \eqref{initial_state_stateprep} and \eqref{state_4_stateprep} we show how the state is prepared using an oracle controlled off the row index.  Therefore each element of $M\ket{b}$ is computed independently allowing one to compute $M$ for the desired row with Eq. \eqref{eq:indepspai}.  This adds a constant overhead to the complexity of calculating $\ket{b}$ alone.

The condition number of the preconditioned matrix can be shown to be constrained to lie in a circle of radius $\sqrt{d}\epsilon_{pre}$, where $\epsilon_{pre} > ||A m_k - e_k||$ is the largest residual of any preconditioned matrix row from the identity \cite{doi:10.1137/S1064827594276552}. If $\sqrt{d} \epsilon_{pre} < 1$ then the spectral condition number satisfies the inequality
\begin{equation}
\label{eq:spectralconditionnumber}
\kappa \equiv \left| \frac{\lambda_{max}}{\lambda_{min}}\right| \le \frac{1 + \sqrt{d}\epsilon_{pre}}{1-\sqrt{d}\epsilon_{pre}}.
\end{equation}

We now show how our algorithm can achieve exponential speedup over the best classical algorithm. On a classical computer the runtime is dominated by the linear systems solving operation that requires many matrix vector products.  The best sparse-matrix solving algorithm, conjugate gradient, is $O[N d \kappa \log(1/\epsilon)]$, where $d$ is the number of non-zero entries per row and $\kappa$ is the condition number of the matrix, while $\epsilon$ is the desired precision of the calculation \cite{saad2003iterative}.

The quantum algorithm requires $O(d^2)$ oracle queries and $O(d^3)$ computational steps to create $M\ket{b}$ and $O(1)$ queries to create $\ket{R}$. Estimation of $\sin^2\phi_b$ and $\sin^2\phi_r$ requires $O(1/\epsilon)$ iterations to estimate to accuracy $\epsilon$ with \ac{AE}.   The \ac{QLSA} requires Hamiltonian simulation to invert $A$.  Berry \textit{et al.} \cite{springerlink:10.1007/s00220-006-0150-x} show that when using the Suzuki higher order integrator method \cite{Suzuki1990319}, this step requires $N_{\textnormal{exp}} \leq 2 m^2 \tau \exp(2\sqrt{\ln 5 \ln(m \tau/\epsilon)})$ exponential operator applications, where $m$ is the number of sub-matrices needed to decompose the sparse matrix $A$ into 1-sparse form ($m=6d^2$ using the decomposition technique in Ref. \cite{springerlink:10.1007/s00220-006-0150-x}, $d$ is the sparsity of $A$, where sparsity is defined as the maximum number of non-zero elements per row), and $\tau=\kappa||A||/\epsilon$.  The overhead to estimating the preconditioner varies depending on which technique one uses for estimating the sparsity pattern.  As an example, if one uses an a priori sparsity pattern \cite{doi:10.1137/S106482759833913X} then one must simply solve a small $O(n\times d)$ linear system, which takes $O(d^3)$ operations and $O(d^2)$ $A$ matrix oracle queries.  For the algorithm to be accurate to within $\epsilon$, Harrow \textit{et al.} \cite{PhysRevLett.103.150502} showed that $\tau=O(\kappa/\epsilon)$.  Since we estimate $\phi_x$ using \ac{AE}, multiple applications of Hamiltonian simulation with different times are required.  Thus, to estimate $\sin^2\phi_x$ as well as $P_{1110}$ and $P_{1111}$   to accuracy $\epsilon$ takes $\tilde{O}(d^7\kappa\log N/\epsilon^2)$ where the tilde indicates that we are neglecting more slowly growing terms in the exponent of $N_{\textnormal{exp}}$.  Our implementation is quadratically better in $\kappa$ than in the original \ac{QLSA} due to our removal of the post-selection step.

Combining all steps, the overall quantum algorithm has $\tilde{O}(d^7\kappa\log N/\epsilon^2)$ complexity.  When the \ac{SPAI} can compute a preconditioner efficiently this algorithm provides exponential speedup over the best classical algorithm, since the condition number is bounded by Eq. \eqref{eq:spectralconditionnumber}.  The \ac{SPAI} preconditioner is known to be applicable to a wide class of problems \cite{doi:10.1137/S1064827594276552,doi:10.1137/S106482759833913X, benzi2002preconditioning, 1710907,ping2009factorized}, greatly expanding the number of applications that can achieve exponential speedup over a classical solution method.

To demonstrate the algorithm's applicability, we now show how it can be used to calculate the electromagnetic scattering cross section of an arbitrary target using the \ac{FEM} \cite{jin02}.  Calculation of the scattering cross section is routinely used in the electromagnetics modeling community to characterize detectability by radar.  In particular, the calculations are used to drive design considerations of low-observable (stealth) objects. The \ac{FEM} approach to solving an electromagnetic scattering problem is to break up the computational domain into small volume elements and apply boundary conditions at neighboring  elements.  This allows one to cast the solution of Maxwell's equations into a linear system $\vec{A} \vec{x} = \vec{b}$.  

The matrix $\vec{A}$ is constructed from a discretization of Maxwell's equation together with appropriate boundary conditions due to the scattering object under consideration.  The vector $\vec{b}$ consists of the known electric field components on the scattering boundary.  The matrix $\vec{A}$ and vector $\vec{b}$, which contain information about the scattering object, can be efficiently derived from the components of a matrix $\vec{F}$ that is dependent only upon the form of the discretization chosen to break up the computational domain (see e.g. Ref. \cite{jin02} and the included supplementary material) together with boundary conditions that include the scattering geometry.  Edge basis vectors \cite{chatterjee93}, denoted as $\vec{N}_i$, are highly popular for electromagnetic scattering applications. They give a form of $\vec{F}$ as
\begin{align}
\label{eq:Amatrixeq}
F_{lj} & = \int_V \left[ (\nabla \times \boldsymbol{N}_l ) \cdot (\nabla \times \boldsymbol{N}_j) - k^2 \boldsymbol{N}_l \cdot \boldsymbol{N}_j \right]dV \\ \nonumber
& + i k \int_S (\boldsymbol{N}_l)_t \cdot (\boldsymbol{N}_j)_t dS,
\end{align}
where $V$ is the volume of the computational region, $S$ is the outer surface of the computational region, $k$ is the electric field wavenumber, the subscript $t$ denotes the tangential component, and the indices $l$ and $j$ denote the numbering of all the edges contained in the volume $V$.  The surface integral is an absorbing term used to prevent reflections off the artificial computational boundary.  On the inner scattering surface the correct boundary condition for the scattered field on metallic scatterers is $\hat{\vec{n}} \times \vec{E} = -\hat{\vec{n}} \times \vec{E}^{(i)}$, where $\vec{E}^{(i)}$ is the incident field, $\vec{E}$ is the scattered field, and $\hat{\vec{n}}$ is the unit vector normal to the surface is applied.

Using the edge basis expansion, the far-field radiation in direction $\vec{s}$ is
\begin{equation}
\vec{E}(\vec{s}) \cdot \uvec{p} = \frac{e^{-iks}}{4 \pi s} \sum_{k} R_k(\uvec{s}) x_k,
\end{equation}
where $\uvec{p}$ is the radar polarization (with $\uvec{p} \cdot \uvec{s} = 0$) and
\begin{align}
\label{eq:rvector}
R_k(\uvec{s}, \uvec{p}) & = \uvec{p} \cdot \int_{S} \uvec{s} \times \big\{ \uvec{s} \times \left[ (\nabla \times \vec{N}_k) \times \uvec{n} \right] \\ \nonumber
& + i k \vec{N}_k \times \uvec{n} \big\} e^{i k \uvec{s} \cdot \vec{r}} \, dS,
\end{align}
where the index $k$ here is the global edge index.  The radar scattering cross-section (RCS) in the direction $\uvec{s}$ is given by
\begin{equation}
\text{RCS} = \lim_{s \to \infty} 4 \pi s^2 |\vec{E}(\vec{s}) \cdot \uvec{p}|^2 = \frac{1}{4\pi}|\vec{R} \cdot \vec{x}|^2,
\end{equation}
or simply the dot product of $\vec{R}$ with the solution $\vec{x}$, where we have assumed an incident plane wave with unit electric field amplitude without loss of generality.

The edge basis elements can take a simple functional form, which allows one to analytically evaluate the integrals in Eqs. \eqref{eq:Amatrixeq} and \eqref{eq:rvector}.  This allows for efficient computation of the matrix and vector elements, a requirement for the quantum algorithm.  Because of the local nature of the finite element expansion, the volume and surface integrals extend only over the region encompassed by the finite element.  As a result $\vec{A}$ is highly sparse, allowing an efficient decomposition into a 1-sparse form \cite{springerlink:10.1007/s00220-006-0150-x}, also necessary for the quantum algorithm.

To obtain the cross section using the quantum algorithm one uses the oracles just presented to create the $A$ matrix and $\ket{b}$ and $\ket{R}$ state vectors.  Then one must restore units to the normalized output received from the quantum algorithm.  Doing so yields the following equation for the cross section in terms of outputs from the quantum computation
\begin{equation}
\label{eq:overlap_units3}
\text{RCS} = \frac{1}{4\pi}\frac{N^2\sin^2\phi_b \sin^2\phi_r}{C_b^2C_r^2\sin^2\phi_x}(P_{1110}-P_{1111}),
\end{equation} 
where $C_b = 1/\text{max}(\vec{b})$ and $C_r = 1/\text{max}(\vec{R})$ are known parameters.  Thus to compute the cross section, we estimate each $\sin^2{\phi_{(b,x,r)}}$ term as well as the $P_{1110}$ and $P_{1111}$ terms independently using \ac{AE}.  

Finally we remark on the efficiency of the scattering cross section calculation.  With no preconditioning, finite element condition numbers scale as $N^{2/n}$ \cite{Brenner2008,bank1989}, where $n$ is the number of dimensions of the problem, implying that even in the most general case our algorithm scales better than its classical counterpart for a three-dimensional finite element problems.  However, by applying the quantum preconditioner, the eigenvalues of the finite element matrix can be bounded achieving exponential speedup, since the \ac{FEM} admits an efficient \ac{SPAI} \cite{1710907, ping2009factorized}.

We have demonstrated a quantum algorithm that generalizes the \ac{QLSA} to solve arbitrary linear systems.  We show how simple ancilla measurements can efficiently calculate many useful quantities of interest from the exponentially large solution space.  Additionally, we have greatly expanded the class of problems that can be solved with exponential speedup, by incorporating matrix preconditioning into the quantum algorithm.  To demonstrate its functionality we showed how one could use it to solve an electromagnetic scattering problem using the finite element method and estimate the scattering cross section.  We show that this can be done in a time exponentially faster than the best classical algorithm.  This opens up the potential for quantum computing to be applied to a broad class of problems of practical interest to the computational physics community.

\acknowledgments{This project was supported by the Intelligence Advanced Research Projects Activity via Department of Interior National Business Center contract numbers N00024-03-D-6606 and 2012-12050800010, with additional support provided by a Stuart S. Janney Fellowship from the Applied Physics Laboratory. The U.S. Government is authorized to reproduce and distribute reprints for Governmental purposes notwithstanding any copyright annotation thereon. The views and conclusions contained herein are those of the authors and should not be interpreted as necessarily representing the official policies or endorsements, either expressed or implied, of IARPA, DoI/NBC, or the U.S. Government. Many thanks to Joan Hoffmann and Nathan Wiebe for helpful comments and discussions.}


\begin{thebibliography}{25}%
\makeatletter
\providecommand \@ifxundefined [1]{%
 \@ifx{#1\undefined}
}%
\providecommand \@ifnum [1]{%
 \ifnum #1\expandafter \@firstoftwo
 \else \expandafter \@secondoftwo
 \fi
}%
\providecommand \@ifx [1]{%
 \ifx #1\expandafter \@firstoftwo
 \else \expandafter \@secondoftwo
 \fi
}%
\providecommand \natexlab [1]{#1}%
\providecommand \enquote  [1]{``#1''}%
\providecommand \bibnamefont  [1]{#1}%
\providecommand \bibfnamefont [1]{#1}%
\providecommand \citenamefont [1]{#1}%
\providecommand \href@noop [0]{\@secondoftwo}%
\providecommand \href [0]{\begingroup \@sanitize@url \@href}%
\providecommand \@href[1]{\@@startlink{#1}\@@href}%
\providecommand \@@href[1]{\endgroup#1\@@endlink}%
\providecommand \@sanitize@url [0]{\catcode `\\12\catcode `\$12\catcode
  `\&12\catcode `\#12\catcode `\^12\catcode `\_12\catcode `\%12\relax}%
\providecommand \@@startlink[1]{}%
\providecommand \@@endlink[0]{}%
\providecommand \url  [0]{\begingroup\@sanitize@url \@url }%
\providecommand \@url [1]{\endgroup\@href {#1}{\urlprefix }}%
\providecommand \urlprefix  [0]{URL }%
\providecommand \Eprint [0]{\href }%
\@ifxundefined \urlstyle {%
  \providecommand \doi  [0]{\begingroup \@sanitize@url \@doi}%
  \providecommand \@doi [1]{\endgroup \@@startlink {\doibase
  #1}doi:\discretionary {}{}{}#1\@@endlink }%
}{%
  \providecommand \doi  [0]{doi:\discretionary{}{}{}\begingroup
  \urlstyle{rm}\Url }%
}%
\providecommand \doibase [0]{http://dx.doi.org/}%
\providecommand \Doi [0]{\begingroup \@sanitize@url \@Doi }%
\providecommand \@Doi  [1]{\endgroup\@@startlink{\doibase#1}\@@Doi}%
\providecommand \@@Doi [1]{#1\@@endlink}%
\providecommand \selectlanguage [0]{\@gobble}%
\providecommand \bibinfo  [0]{\@secondoftwo}%
\providecommand \bibfield  [0]{\@secondoftwo}%
\providecommand \translation [1]{[#1]}%
\providecommand \BibitemOpen [0]{}%
\providecommand \bibitemStop [0]{}%
\providecommand \bibitemNoStop [0]{.\EOS\space}%
\providecommand \EOS [0]{\spacefactor3000\relax}%
\providecommand \BibitemShut  [1]{\csname bibitem#1\endcsname}%
\bibitem [{\citenamefont {Feynman}(1982)}]{springerlink:10.1007/BF02650179}%
  \BibitemOpen
  \bibfield  {author} {\bibinfo {author} {\bibfnamefont {R.}~\bibnamefont
  {Feynman}},\ }\Doi {10.1007/BF02650179} {\bibfield  {journal} {\bibinfo
  {journal} {International Journal of Theoretical Physics},\ }\textbf {\bibinfo
  {volume} {21}},\ \bibinfo {pages} {467} (\bibinfo {year} {1982})}\BibitemShut
  {NoStop}%
\bibitem [{\citenamefont {Shor}(1994)}]{Shor1994}%
  \BibitemOpen
  \bibfield  {author} {\bibinfo {author} {\bibfnamefont {P.}~\bibnamefont
  {Shor}},\ }in\ \Doi {10.1109/SFCS.1994.365700} {\emph {\bibinfo {booktitle}
  {Foundations of Computer Science, 1994 Proceedings., 35th Annual Symposium
  on}}}\ (\bibinfo {year} {1994})\ pp.\ \bibinfo {pages} {124
  --134}\BibitemShut {NoStop}%
\bibitem [{\citenamefont {Harrow}\ \emph {et~al.}(2009)\citenamefont {Harrow},
  \citenamefont {Hassidim},\ and\ \citenamefont
  {Lloyd}}]{PhysRevLett.103.150502}%
  \BibitemOpen
  \bibfield  {author} {\bibinfo {author} {\bibfnamefont {A.~W.}\ \bibnamefont
  {Harrow}}, \bibinfo {author} {\bibfnamefont {A.}~\bibnamefont {Hassidim}}, \
  and\ \bibinfo {author} {\bibfnamefont {S.}~\bibnamefont {Lloyd}},\ }\Doi
  {10.1103/PhysRevLett.103.150502} {\bibfield  {journal} {\bibinfo  {journal}
  {Phys. Rev. Lett.},\ }\textbf {\bibinfo {volume} {103}},\ \bibinfo {pages}
  {150502} (\bibinfo {year} {2009})}\BibitemShut {NoStop}%
\bibitem [{\citenamefont {Grover}\ and\ \citenamefont
  {Rudolph}(2002)}]{Grover02}%
  \BibitemOpen
  \bibfield  {author} {\bibinfo {author} {\bibfnamefont {L.}~\bibnamefont
  {Grover}}\ and\ \bibinfo {author} {\bibfnamefont {T.}~\bibnamefont
  {Rudolph}},\ }\href {http://arxiv.org/abs/quant-ph/0208112} {\bibfield
  {journal} {\bibinfo  {journal} {arXiv:quant-ph/0208112v1}} (\bibinfo {year}
  {2002})}\BibitemShut {NoStop}%
\bibitem [{\citenamefont {Aharonov}\ and\ \citenamefont
  {Ta-Shma}(2003){\natexlab{a}}}]{Aharonov:2003:AQS:780542.780546}%
  \BibitemOpen
  \bibfield  {author} {\bibinfo {author} {\bibfnamefont {D.}~\bibnamefont
  {Aharonov}}\ and\ \bibinfo {author} {\bibfnamefont {A.}~\bibnamefont
  {Ta-Shma}},\ }in\ \Doi {http://doi.acm.org/10.1145/780542.780546} {\emph
  {\bibinfo {booktitle} {Proceedings of the thirty-fifth annual ACM symposium
  on Theory of computing}}},\ \bibinfo {series and number} {STOC '03}\
  (\bibinfo  {publisher} {ACM},\ \bibinfo {address} {New York, NY, USA},\
  \bibinfo {year} {2003})\ pp.\ \bibinfo {pages} {20--29}\BibitemShut {NoStop}%
\bibitem [{\citenamefont {Ward}\ \emph {et~al.}(2009)\citenamefont {Ward},
  \citenamefont {Kassal},\ and\ \citenamefont {Aspuru-Guzik}}]{ward:194105}%
  \BibitemOpen
  \bibfield  {author} {\bibinfo {author} {\bibfnamefont {N.~J.}\ \bibnamefont
  {Ward}}, \bibinfo {author} {\bibfnamefont {I.}~\bibnamefont {Kassal}}, \ and\
  \bibinfo {author} {\bibfnamefont {A.}~\bibnamefont {Aspuru-Guzik}},\ }\Doi
  {10.1063/1.3115177} {\bibfield  {journal} {\bibinfo  {journal} {The Journal
  of Chemical Physics},\ }\textbf {\bibinfo {volume} {130}},\ \bibinfo {eid}
  {194105} (\bibinfo {year} {2009})}\BibitemShut {NoStop}%
\bibitem [{\citenamefont {Soklakov}\ and\ \citenamefont
  {Schack}(2006)}]{PhysRevA.73.012307}%
  \BibitemOpen
  \bibfield  {author} {\bibinfo {author} {\bibfnamefont {A.~N.}\ \bibnamefont
  {Soklakov}}\ and\ \bibinfo {author} {\bibfnamefont {R.}~\bibnamefont
  {Schack}},\ }\Doi {10.1103/PhysRevA.73.012307} {\bibfield  {journal}
  {\bibinfo  {journal} {Phys. Rev. A},\ }\textbf {\bibinfo {volume} {73}},\
  \bibinfo {pages} {012307} (\bibinfo {year} {2006})}\BibitemShut {NoStop}%
\bibitem [{\citenamefont {Kaye}\ and\ \citenamefont {Mosca}(2004)}]{Kaye04}%
  \BibitemOpen
  \bibfield  {author} {\bibinfo {author} {\bibfnamefont {P.}~\bibnamefont
  {Kaye}}\ and\ \bibinfo {author} {\bibfnamefont {M.}~\bibnamefont {Mosca}},\
  }\href {http://arxiv.org/abs/quant-ph/0407102} {\bibfield  {journal}
  {\bibinfo  {journal} {arXiv:quant-ph/0407102v1}} (\bibinfo {year}
  {2004})}\BibitemShut {NoStop}%
\bibitem [{\citenamefont {Brassard}\ \emph {et~al.}(2000)\citenamefont
  {Brassard}, \citenamefont {Hoyer}, \citenamefont {Mosca},\ and\ \citenamefont
  {Tapp}}]{Brassard00}%
  \BibitemOpen
  \bibfield  {author} {\bibinfo {author} {\bibfnamefont {G.}~\bibnamefont
  {Brassard}}, \bibinfo {author} {\bibfnamefont {P.}~\bibnamefont {Hoyer}},
  \bibinfo {author} {\bibfnamefont {M.}~\bibnamefont {Mosca}}, \ and\ \bibinfo
  {author} {\bibfnamefont {A.}~\bibnamefont {Tapp}},\ }\href
  {http://arxiv.org/abs/quant-ph/0005055v1} {\bibfield  {journal} {\bibinfo
  {journal} {arXiv.org:quant-ph/0005055}} (\bibinfo {year} {2000})}\BibitemShut
  {NoStop}%
\bibitem [{\citenamefont {Wiebe}\ \emph {et~al.}(2012)\citenamefont {Wiebe},
  \citenamefont {Braun},\ and\ \citenamefont {Lloyd}}]{PhysRevLett.109.050505}%
  \BibitemOpen
  \bibfield  {author} {\bibinfo {author} {\bibfnamefont {N.}~\bibnamefont
  {Wiebe}}, \bibinfo {author} {\bibfnamefont {D.}~\bibnamefont {Braun}}, \ and\
  \bibinfo {author} {\bibfnamefont {S.}~\bibnamefont {Lloyd}},\ }\Doi
  {10.1103/PhysRevLett.109.050505} {\bibfield  {journal} {\bibinfo  {journal}
  {Phys. Rev. Lett.},\ }\textbf {\bibinfo {volume} {109}},\ \bibinfo {pages}
  {050505} (\bibinfo {year} {2012})}\BibitemShut {NoStop}%
\bibitem [{\citenamefont {Brenner}\ and\ \citenamefont
  {Scott}(2008)}]{Brenner2008}%
  \BibitemOpen
  \bibfield  {author} {\bibinfo {author} {\bibfnamefont {S.~C.}\ \bibnamefont
  {Brenner}}\ and\ \bibinfo {author} {\bibfnamefont {L.~R.}\ \bibnamefont
  {Scott}},\ }\href@noop {} {\emph {\bibinfo {title} {{The Mathematical Theory
  of Finite Element Methods}}}},\ \bibinfo {edition} {3rd}\ ed.\ (\bibinfo
  {publisher} {Springer Verlag, New York},\ \bibinfo {year} {2008})\ Chap.\
  \bibinfo {chapter} {9.6}\BibitemShut {NoStop}%
\bibitem [{\citenamefont {Bank}\ and\ \citenamefont {Scott}(1989)}]{bank1989}%
  \BibitemOpen
  \bibfield  {author} {\bibinfo {author} {\bibfnamefont {R.~E.}\ \bibnamefont
  {Bank}}\ and\ \bibinfo {author} {\bibfnamefont {L.~R.}\ \bibnamefont
  {Scott}},\ }\Doi {10.1137/0726080} {\bibfield  {journal} {\bibinfo  {journal}
  {SIAM Journal on Numerical Analysis},\ }\textbf {\bibinfo {volume} {26}},\
  \bibinfo {pages} {1383} (\bibinfo {year} {1989})}\BibitemShut {NoStop}%
\bibitem [{\citenamefont {Benzi}(2002)}]{benzi2002preconditioning}%
  \BibitemOpen
  \bibfield  {author} {\bibinfo {author} {\bibfnamefont {M.}~\bibnamefont
  {Benzi}},\ }\href@noop {} {\bibfield  {journal} {\bibinfo  {journal} {Journal
  of Computational Physics},\ }\textbf {\bibinfo {volume} {182}},\ \bibinfo
  {pages} {418} (\bibinfo {year} {2002})}\BibitemShut {NoStop}%
\bibitem [{\citenamefont {Grote}\ and\ \citenamefont
  {Huckle}(1997)}]{doi:10.1137/S1064827594276552}%
  \BibitemOpen
  \bibfield  {author} {\bibinfo {author} {\bibfnamefont {M.}~\bibnamefont
  {Grote}}\ and\ \bibinfo {author} {\bibfnamefont {T.}~\bibnamefont {Huckle}},\
  }\Doi {10.1137/S1064827594276552} {\bibfield  {journal} {\bibinfo  {journal}
  {SIAM Journal on Scientific Computing},\ }\textbf {\bibinfo {volume} {18}},\
  \bibinfo {pages} {838} (\bibinfo {year} {1997})}\BibitemShut {NoStop}%
\bibitem [{\citenamefont {Chow}(2000)}]{doi:10.1137/S106482759833913X}%
  \BibitemOpen
  \bibfield  {author} {\bibinfo {author} {\bibfnamefont {E.}~\bibnamefont
  {Chow}},\ }\Doi {10.1137/S106482759833913X} {\bibfield  {journal} {\bibinfo
  {journal} {SIAM Journal on Scientific Computing},\ }\textbf {\bibinfo
  {volume} {21}},\ \bibinfo {pages} {1804} (\bibinfo {year}
  {2000})}\BibitemShut {NoStop}%
\bibitem [{\citenamefont {Aharonov}\ and\ \citenamefont
  {Ta-Shma}(2003){\natexlab{b}}}]{aharonov2003adiabatic}%
  \BibitemOpen
  \bibfield  {author} {\bibinfo {author} {\bibfnamefont {D.}~\bibnamefont
  {Aharonov}}\ and\ \bibinfo {author} {\bibfnamefont {A.}~\bibnamefont
  {Ta-Shma}},\ }in\ \href@noop {} {\emph {\bibinfo {booktitle} {Proceedings of
  the thirty-fifth annual ACM symposium on Theory of computing}}}\ (\bibinfo
  {organization} {ACM},\ \bibinfo {year} {2003})\ pp.\ \bibinfo {pages}
  {20--29}\BibitemShut {NoStop}%
\bibitem [{\citenamefont {Childs}\ \emph {et~al.}(2003)\citenamefont {Childs},
  \citenamefont {Cleve}, \citenamefont {Deotto}, \citenamefont {Farhi},
  \citenamefont {Gutmann},\ and\ \citenamefont
  {Spielman}}]{Childs:2003:EAS:780542.780552}%
  \BibitemOpen
  \bibfield  {author} {\bibinfo {author} {\bibfnamefont {A.~M.}\ \bibnamefont
  {Childs}}, \bibinfo {author} {\bibfnamefont {R.}~\bibnamefont {Cleve}},
  \bibinfo {author} {\bibfnamefont {E.}~\bibnamefont {Deotto}}, \bibinfo
  {author} {\bibfnamefont {E.}~\bibnamefont {Farhi}}, \bibinfo {author}
  {\bibfnamefont {S.}~\bibnamefont {Gutmann}}, \ and\ \bibinfo {author}
  {\bibfnamefont {D.~A.}\ \bibnamefont {Spielman}},\ }in\ \Doi
  {10.1145/780542.780552} {\emph {\bibinfo {booktitle} {Proceedings of the
  thirty-fifth annual ACM symposium on Theory of computing}}},\ \bibinfo
  {series and number} {STOC '03}\ (\bibinfo  {publisher} {ACM},\ \bibinfo
  {address} {New York, NY, USA},\ \bibinfo {year} {2003})\ pp.\ \bibinfo
  {pages} {59--68},\ ISBN \bibinfo {isbn} {1-58113-674-9}\BibitemShut {NoStop}%
\bibitem [{\citenamefont {Wiebe}\ \emph {et~al.}(2011)\citenamefont {Wiebe},
  \citenamefont {Berry}, \citenamefont {Hoyer},\ and\ \citenamefont
  {Sanders}}]{1751-8121-44-44-445308}%
  \BibitemOpen
  \bibfield  {author} {\bibinfo {author} {\bibfnamefont {N.}~\bibnamefont
  {Wiebe}}, \bibinfo {author} {\bibfnamefont {D.~W.}\ \bibnamefont {Berry}},
  \bibinfo {author} {\bibfnamefont {P.}~\bibnamefont {Hoyer}}, \ and\ \bibinfo
  {author} {\bibfnamefont {B.~C.}\ \bibnamefont {Sanders}},\ }\href
  {http://stacks.iop.org/1751-8121/44/i=44/a=445308} {\bibfield  {journal}
  {\bibinfo  {journal} {Journal of Physics A: Mathematical and Theoretical},\
  }\textbf {\bibinfo {volume} {44}},\ \bibinfo {pages} {445308} (\bibinfo
  {year} {2011})}\BibitemShut {NoStop}%
\bibitem [{\citenamefont {Saad}(2003)}]{saad2003iterative}%
  \BibitemOpen
  \bibfield  {author} {\bibinfo {author} {\bibfnamefont {Y.}~\bibnamefont
  {Saad}},\ }\href@noop {} {\emph {\bibinfo {title} {Iterative methods for
  sparse linear systems}}}\ (\bibinfo  {publisher} {Society for Industrial and
  Applied Mathematics},\ \bibinfo {year} {2003})\BibitemShut {NoStop}%
\bibitem [{\citenamefont {Berry}\ \emph {et~al.}(2007)\citenamefont {Berry},
  \citenamefont {Ahokas}, \citenamefont {Cleve},\ and\ \citenamefont
  {Sanders}}]{springerlink:10.1007/s00220-006-0150-x}%
  \BibitemOpen
  \bibfield  {author} {\bibinfo {author} {\bibfnamefont {D.}~\bibnamefont
  {Berry}}, \bibinfo {author} {\bibfnamefont {G.}~\bibnamefont {Ahokas}},
  \bibinfo {author} {\bibfnamefont {R.}~\bibnamefont {Cleve}}, \ and\ \bibinfo
  {author} {\bibfnamefont {B.}~\bibnamefont {Sanders}},\ }\Doi
  {10.1007/s00220-006-0150-x} {\bibfield  {journal} {\bibinfo  {journal}
  {Communications in Mathematical Physics},\ }\textbf {\bibinfo {volume}
  {270}},\ \bibinfo {pages} {359} (\bibinfo {year} {2007})}\BibitemShut
  {NoStop}%
\bibitem [{\citenamefont {Suzuki}(1990)}]{Suzuki1990319}%
  \BibitemOpen
  \bibfield  {author} {\bibinfo {author} {\bibfnamefont {M.}~\bibnamefont
  {Suzuki}},\ }\Doi {10.1016/0375-9601(90)90962-N} {\bibfield  {journal}
  {\bibinfo  {journal} {Physics Letters A},\ }\textbf {\bibinfo {volume}
  {146}},\ \bibinfo {pages} {319 } (\bibinfo {year} {1990})}\BibitemShut
  {NoStop}%
\bibitem [{\citenamefont {Li}\ \emph {et~al.}(2006)\citenamefont {Li},
  \citenamefont {Rui},\ and\ \citenamefont {Chen}}]{1710907}%
  \BibitemOpen
  \bibfield  {author} {\bibinfo {author} {\bibfnamefont {S.}~\bibnamefont
  {Li}}, \bibinfo {author} {\bibfnamefont {P.}~\bibnamefont {Rui}}, \ and\
  \bibinfo {author} {\bibfnamefont {R.}~\bibnamefont {Chen}},\ }in\ \Doi
  {10.1109/APS.2006.1710907} {\emph {\bibinfo {booktitle} {Antennas and
  Propagation Society International Symposium 2006, IEEE}}}\ (\bibinfo {year}
  {2006})\ pp.\ \bibinfo {pages} {1765--1768}\BibitemShut {NoStop}%
\bibitem [{\citenamefont {Ping}\ and\ \citenamefont
  {Cui}(2009)}]{ping2009factorized}%
  \BibitemOpen
  \bibfield  {author} {\bibinfo {author} {\bibfnamefont {X.~W.}\ \bibnamefont
  {Ping}}\ and\ \bibinfo {author} {\bibfnamefont {T.-J.}\ \bibnamefont {Cui}},\
  }\href@noop {} {\bibfield  {journal} {\bibinfo  {journal} {Progress In
  Electromagnetics Research},\ }\textbf {\bibinfo {volume} {98}},\ \bibinfo
  {pages} {15} (\bibinfo {year} {2009})}\BibitemShut {NoStop}%
\bibitem [{\citenamefont {Jin}(2002)}]{jin02}%
  \BibitemOpen
  \bibfield  {author} {\bibinfo {author} {\bibfnamefont {J.}~\bibnamefont
  {Jin}},\ }\href@noop {} {\emph {\bibinfo {title} {{The Finite Element Method
  in Electromagnetics}}}},\ \bibinfo {edition} {2nd}\ ed.\ (\bibinfo
  {publisher} {John Wiley and Sons, Inc.},\ \bibinfo {year} {2002})\BibitemShut
  {NoStop}%
\bibitem [{\citenamefont {Chatterjee}\ \emph {et~al.}(1993)\citenamefont
  {Chatterjee}, \citenamefont {Jin},\ and\ \citenamefont
  {Volakis}}]{chatterjee93}%
  \BibitemOpen
  \bibfield  {author} {\bibinfo {author} {\bibfnamefont {A.}~\bibnamefont
  {Chatterjee}}, \bibinfo {author} {\bibfnamefont {J.}~\bibnamefont {Jin}}, \
  and\ \bibinfo {author} {\bibfnamefont {J.}~\bibnamefont {Volakis}},\ }\Doi
  {10.1109/8.214614} {\bibfield  {journal} {\bibinfo  {journal} {Antennas and
  Propagation, IEEE Transactions on},\ }\textbf {\bibinfo {volume} {41}},\
  \bibinfo {pages} {221 } (\bibinfo {year} {1993})}\BibitemShut {NoStop}%
\end{thebibliography}
\end{document}